# Controlling Neuronal Noise Using Chaos Control

David J. Christini and James J. Collins

*NeuroMuscular Research Center and Department of Biomedical Engineering*

*Boston University, 44 Cummington Street, Boston, Massachusetts 02215*

(March 15, 1995)

Chaos control techniques have been applied to a wide variety of experimental systems, including magneto-elastic ribbons [1], lasers [2], chemical reactions [3], arrhythmic cardiac tissue [4], and spontaneously bursting neuronal networks [5]. An underlying assumption in all of these studies is that the system being controlled is chaotic. However, the identification of chaos in experimental systems, particularly physiological systems, is a difficult and often misleading task [6–9]. Here we demonstrate that the chaos criteria used in a recent study [5] can falsely classify a noise-driven, non-chaotic neuronal model as being chaotic. We apply chaos control, periodic pacing, and anticontrol to the non-chaotic model and obtain results which are similar to those reported for apparently chaotic, *in vitro* neuronal networks [5]. We also obtain similar results when we apply chaos control to a simple stochastic system. These novel findings challenge the claim that the aforementioned neuronal networks [5] were chaotic and suggest that chaos control techniques can be applied to a wider range of experimental systems than previously thought.

Schiff *et al.* [5] studied the firing behavior of neuronal networks in hippocampal slices of rat brain. They generated noise-like (possibly chaotic) burst-firing activity in these networks by exposing the hippocampal slices to $K^+$-enriched artificial cerebrospinal fluid. As a simple analogue to this system, we considered the firing behavior of a noise-driven neuron. Specifically, we implemented the FitzHugh-Nagumo neuronal model (see Fig. 1 caption for details). In the present case, the model neuron was driven by both tonic and noisy inputs. The system parameter values were chosen such that the model neuron fired periodically in the absence of noisy inputs. Phase-plane analysis showed that the additive noise simply caused the firing period to fluctuate about its mean value. The periodic orbit was structurally preserved. There were no global bifurcations; thus, the additive noise did not induce chaos.

To evaluate the presence of chaos in the aforementioned neuronal networks, Schiff *et al.* [5] analyzed the first-return maps of the burst interspike intervals (ISI) for different hippocampal-slice preparations. According to their criteria, a system could be considered chaotic if its ISI time series contained recurrent sequences which approached an unstable periodic flip-saddle fixed point (in the first-return map) along a stable direction (manifold) and departed from it in an exponential fashion along a locally-linear unstable manifold. The ISI first-return maps for the *in vitro* neuronal networks of Schiff *et al.* [5] satisfied the above chaos criteria. However, we found that the ISI time series from our noise-driven, non-chaotic neuronal model also satisfied these criteria (Fig. 1), i.e., the model's time series contained recurrent sequences which approached and exponentially diverged from an apparent unstable periodic fixed point in the first-return map (Fig. 1*b*). (It should also be noted that the plots in Fig. 1 are similar in structure to those in Fig. 2 in ref. [5].) We use the term "apparent" to describe the candidate fixed points in our model's output because the aforementioned phase-plane analysis showed that for the parameter values used, the model does not have any unstable periodic fixed points. To confirm this finding, we generated ten 5000-ISI time series from the noise-driven model neuron. For each ISI time series, we then generated a set of ten randomly shuffled surrogate data sets. We found that the probability of finding candidate unstable periodic fixed points in the original time series was not statistically significantly different from the probability of finding such points in the respective surrogate data sets (*p*-values ranged from 0.37-0.90, mean 0.61). These results, together with those in Fig. 1, demonstrate that apparent unstable periodic fixed points can arise simply by chance in our noise-driven, non-chaotic neuronal model, and therefore, the chaos criteria used by Schiff *et al.* [5] are not sufficient for the definitive identification of deterministic chaos.

The original chaos control technique developed by Ott, Grebogi, and Yorke [10] (OGY) is based on the fact that there are an infinite number of unstable periodic orbits embedded within a chaotic attractor. With this approach, a chaotic system is stabilized about one of these periodic orbits by making small time-dependent perturbations to an accessible systemwide parameter such that the system's trajectory is directed toward the stable manifold of the desired unstable orbit. The OGY technique is useful in many situations because it requires no knowledge of the underlying system equations. Recently, the OGY technique was modified so that chaos control could be applied to systems where no systemwide parameters are readily available for manipulation. This modified method, which is called proportional perturbation feedback (PPF) control [4], involves the application of perturbations to a system variable. With this approach, the goal is to apply perturbations so as to move the system's state point onto the stable manifold of a desired unstable periodic fixed point. Schiff *et al.* [5] implemented PPF control in *in vitro* neuronal networks by delivering precisely-timed electrical stimuli to their hippocampal-slice preparations.

We applied PPF control to our noise-driven, non-chaotic neuronal model and achieved a level of control success (Fig. 2) which was similar to that obtained by Schiff *et al.* [5] for *in vitro* neuronal networks (see Figs 3 and 4 in ref. [5]). For instance, in both our study and the study of Schiff *et al.* [5], the width of the ISI band (i.e., the amount of ISI fluctuation) was reduced considerably from the pre-control stage to the PPF control stage.

To compare PPF control with simple periodic pacing, we repeatedly stimulated the noise-driven, non-chaotic neuronal model at a constant pulse interval. (The periodic-pacing pulse interval was equal to the value of the apparent unstable periodic fixed point used for PPF control.) Our periodic pacing results (Fig. 2) were similar to those reported by Schiff *et al.* [5] for *in vitro* neuronal networks (see Fig. 3 in ref. [5]). In both our study and the study of Schiff *et al.* [5], periodic pacing produced qualitatively different results from PPF control, e.g., periodic pacing frequently allowed ISIs which were considerably larger than the stimulation interval.

It has been suggested that the underlying existence of low-dimensional chaos in the nervous system may offer the opportunity to desynchronize the periodic behavior typical of epileptic seizures [11]. In line with this hypothesis, Schiff *et al.* [5] showed that a technique called anticontrol could be used to reduce the periodicity of their hippocampal-slice preparations. Anticontrol, which is essentially the inverse of chaos control, increases the aperiodicity of a system by moving its state point away from the unstable periodic fixed point. We applied anticontrol (as described in Fig. 2 caption) to our noise-driven, non-chaotic neuronal model and obtained results (Fig. 2) which were similar to those reported by Schiff *et al.* [5] for *in vitro* neuronal networks (see Fig. 4 in ref. [5]), i.e., anticontrol reduced the ISI periodicity in the model neuron.

With PPF control, a system is controlled, in principle, by exploiting the features of one of its unstable periodic fixed points. It was surprising therefore that PPF control could effectively control our noise-driven, non-chaotic neuronal model, given that the model does not have any unstable periodic fixed points. Fig. 3 clarifies this apparent contradiction. In Fig. 3*a*, it can be seen that in repeated control attempts, the effectiveness of control varied as new apparent unstable periodic fixed points were defined, i.e., larger fixed points were associated with decreased levels of control success (similar to the results shown in Fig. 5*b* in ref. [5]). The differences between the respective control regions can be attributed solely to the quantitative differences between the values of the selected fixed points; the response of the system to PPF control was not qualitatively different for the different control regions. Note that the spread (range) of ISIs below each fixed point was not significantly different from that of the uncontrolled regions. (Similar results can be seen in Fig. 5*b* in ref. [5].) PPF control thus largely served to eliminate ISIs that were larger than the value of the selected fixed point. Similar results could have been obtained with demand pacing, which is a simple, well-known technique [12,13] that is not based on chaos theory and that does not require the determination of stable and unstable manifolds - with demand pacing, stimuli are used to prevent the ISIs of a system from exceeding some pre-determined value.

In order to explore the aforementioned points further, we considered a simple stochastic system (see Fig. 3 caption for details) which models the behavior of our noise-driven model neuron to the lowest order. By design, this stochastic system does not have any unstable periodic fixed points and it is incapable of displaying deterministic chaos. However, as with our noise-driven model neuron, this system does display (by chance) "apparent" unstable periodic fixed points. We applied PPF control to this simple system and obtained results (Fig. 3*b*) which were similar to those described above for our noise-driven, non-chaotic neuronal model (Fig. 3*a*) and those reported by Schiff *et al.* [5] for *in vitro* neuronal networks (see Fig. 5*b* in ref. [5]). This work clearly demonstrates that PPF control can control a stochastic system, one which does not have any unstable periodic fixed points, with a level of success similar to that reported by Schiff *et al.* [5]. These results thereby challenge Schiff *et al.*'s [5] claim of having controlled chaos in their hippocampal-slice preparations. This point is corroborated by a recent study [14] which showed that the majority of the ISI time series from similar hippocampal-slice preparations failed to demonstrate evidence of deterministic structure.

The recent success of chaos control in physiology has led to speculations that these techniques may be clinically useful [4,5,15]. The present findings do not discount that possibility; rather, our work suggests that chaos control techniques can be applied to a wider range of experimental systems, e.g., stochastic systems, than previously thought. Moreover, because PPF control can be applied to both chaotic and non-chaotic systems, the difficult problem of distinguishing between deterministic chaos and noise in physiological systems [6–9] appears to be a non-issue for this application. Clearly, however, the nature of the control success will depend critically upon the characteristics (e.g., the presence of unstable periodic fixed points) of the system to be controlled.

## ACKNOWLEDGMENTS

We thank Dr. Carson Chow for useful discussions. We thank Thomas Imhoff for assistance with the implementation of the FitzHugh-Nagumo model. This work was supported by the National Science Foundation.

FIGURES

FIG. 1. Plots of interspike intervals $ISI_n$ versus the previous interval $ISI_{n-1}$ for the noise-driven, non-chaotic neuronal model without control. The interspike intervals were computed using the method described by Longtin [16]. For all results presented in this paper, we used the FitzHugh-Nagumo (FHN) neuronal model [16] as given by the following equations:

$\epsilon \frac{dv(t)}{dt} = v(t)[v(t) - a][1 - v(t)] - w(t) + V_A + \xi(t)$

$\frac{dw(t)}{dt} = v(t) - w(t) - b$

where $v(t)$ is the voltage variable, $w(t)$ is the recovery variable, $V_A$ is a tonic activation signal of 0.2 V, $\xi(t)$ is Gaussian white noise with zero mean and standard deviation=$6.325 \times 10^{-4}$ V, $\epsilon = 0.005$, $a = 0.5$, and $b = 0.15$. The FHN equations were solved numerically using an algorithm developed by Mannella and Palleschi [17] for stochastic differential equations. (The reported results were robust to the integration step size.) In the absence of additive noise, the model fired regularly with a period of 0.761 s. *a*, First-return map showing eight sequential points (numbered 1-8). Points 2-8 define an apparent flip-saddle unstable manifold because the sequence alternates on either side of the line of identity (where $ISI_n = ISI_{n-1}$) while diverging exponentially away from it along a nearly straight line. The intersection of the line of identity with a straight line fit to points 2-8 defines the location of the apparent unstable periodic fixed point. Point 1 lies on an apparent stable manifold because it is followed by point 2 which lies near the apparent unstable periodic fixed point. *b*, First-return map showing multiple trajectories that did not follow each other in time. The starting point for each sequence, numbered 1 in each, began at spike numbers 50 (circle), 105 (triangle), and 788 (square), out of a total series of 791 spikes. The apparent stable manifold is indicated by arrows pointing towards the apparent unstable periodic fixed point, and the apparent unstable manifold is indicated by arrows pointing away from the apparent unstable periodic fixed point. Note that each sequence starts in a region near the apparent stable manifold. The second point for each sequence lies near the apparent unstable periodic fixed point. Each sequence then departs from the apparent unstable periodic fixed point in exponentially diverging jumps on alternating sides of the line of identity along the apparent unstable manifold.

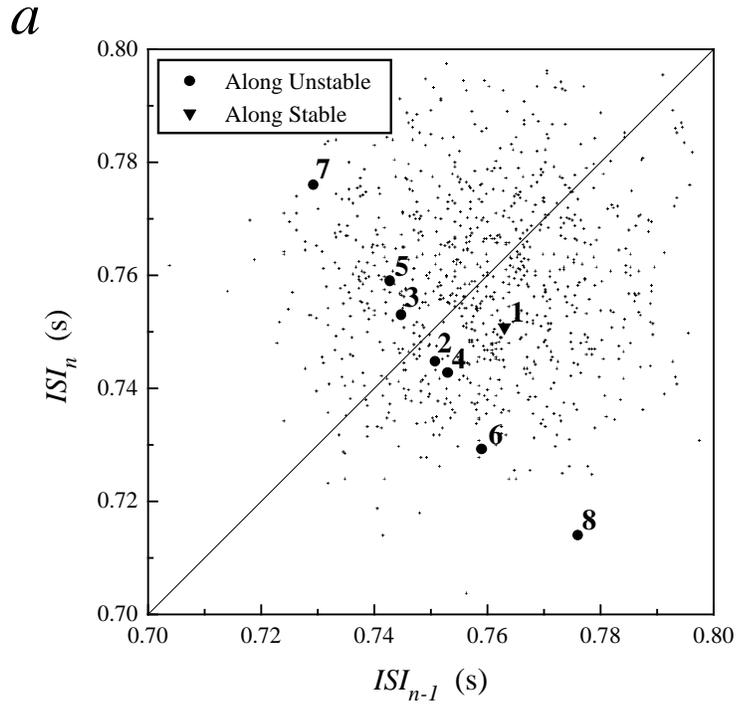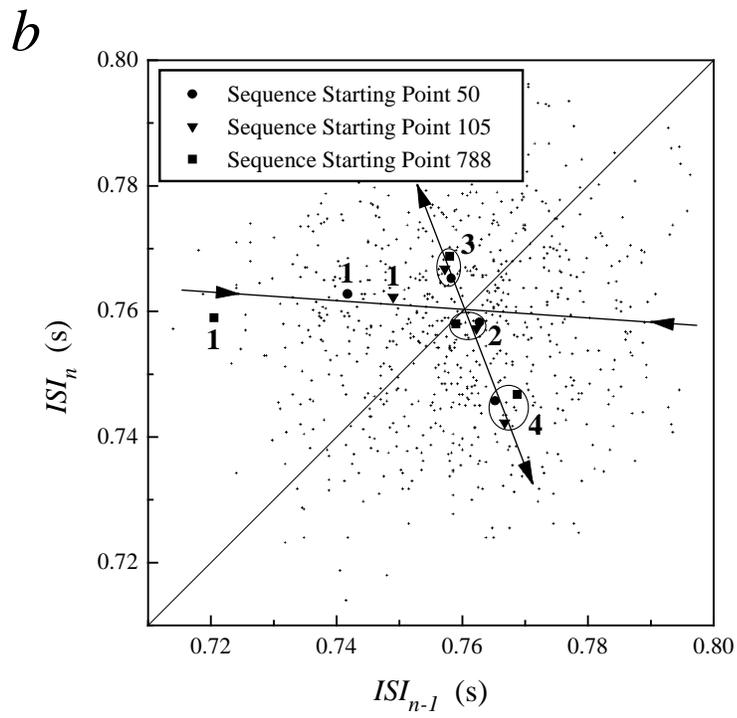

Fig. 1

FIG. 2. Plot showing interspike intervals $ISI_n$ for the noise-driven, non-chaotic neuronal model without control (temporal regions $A$), and with proportional perturbation feedback (PPF) control (region $B$), periodic pacing (region $C$), and anticontrol (region $D$). Interspike interval number is indicated by $n$. The apparent unstable periodic fixed point for PPF control was defined (during the first region $A$) as the intersection of the line of identity (in the first-return map) with the apparent unstable and stable manifolds. The apparent unstable manifold was estimated by a linear least-squares fit of consecutive flip-saddle points which diverged exponentially from the line of identity (as for points 2-4 in Fig. 1$b$), while the apparent stable manifold was defined by points which rapidly converged towards the line of identity. The slope of the apparent unstable manifold (which is equivalent to the rate of divergence) was required to be negative with a magnitude greater than 1, while the slope of the apparent stable manifold (which is equivalent to the rate of convergence) was required to have a magnitude less than 1. The simulated electrical stimuli for all control interventions (e.g., PPF control, periodic pacing, and anticontrol) were $250\mu s$ 5V pulses. These pulses were added to the tonic activation signal in the FHN model (see Fig. 1 caption). PPF control was activated for 200 points (region $B$). Following 100 points without control (second region $A$), periodic pacing was activated for 200 points (region $C$). The value of the PPF apparent unstable periodic fixed point from region $B$ was used as the periodic-pacing pulse interval. During the next 100 points without control (third region $A$), the apparent unstable periodic fixed point and apparent unstable manifold for anticontrol were estimated using the techniques described above for PPF control. The algorithm for anticontrol was the same as the algorithm for PPF control except that ISI state points were forced onto an unstable repellor line instead of the apparent stable manifold. For the unstable repellor line, we used the mirror image of the apparent unstable manifold about a vertical line passing through the apparent unstable periodic fixed point.

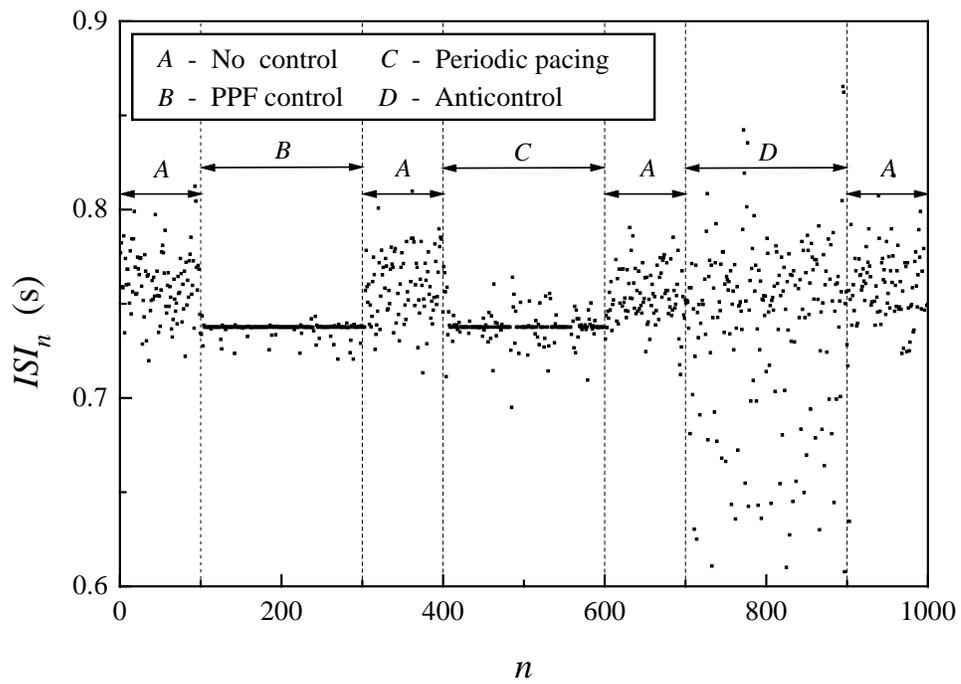

Fig. 2

FIG. 3. *a*, Plot showing interspike intervals $ISI_n$ for the noise-driven, non-chaotic neuronal model without control (temporal regions $A$) and for three distinct periods of proportional perturbation feedback (PPF) control (regions $B$). For each PPF control region, a new apparent unstable periodic fixed point and set of apparent stable and unstable manifolds were determined during the preceding control-free region. It can be seen that the level of control success differed for different fixed points, i.e., the level of control success decreased as the value of the fixed point increased. *b*, Plot showing PPF control (regions $B$) of simulated interspike intervals $ISI_n$ produced by the following simple stochastic system:

$ISI_n = \overline{ISI} + \xi_n$

where $ISI_n$ is a variable which we take to represent the current interspike interval, $\overline{ISI}$ is a constant parameter which represents the mean value (0.761 s) of the interspike interval, and $\xi_n$ is Gaussian white noise with zero mean and standard deviation = 0.02 s. PPF control was implemented as described in *a*. It can be seen that the PPF control results obtained with this simple stochastic system were similar to those obtained with the noise-driven, non-chaotic neuronal model in *a*.

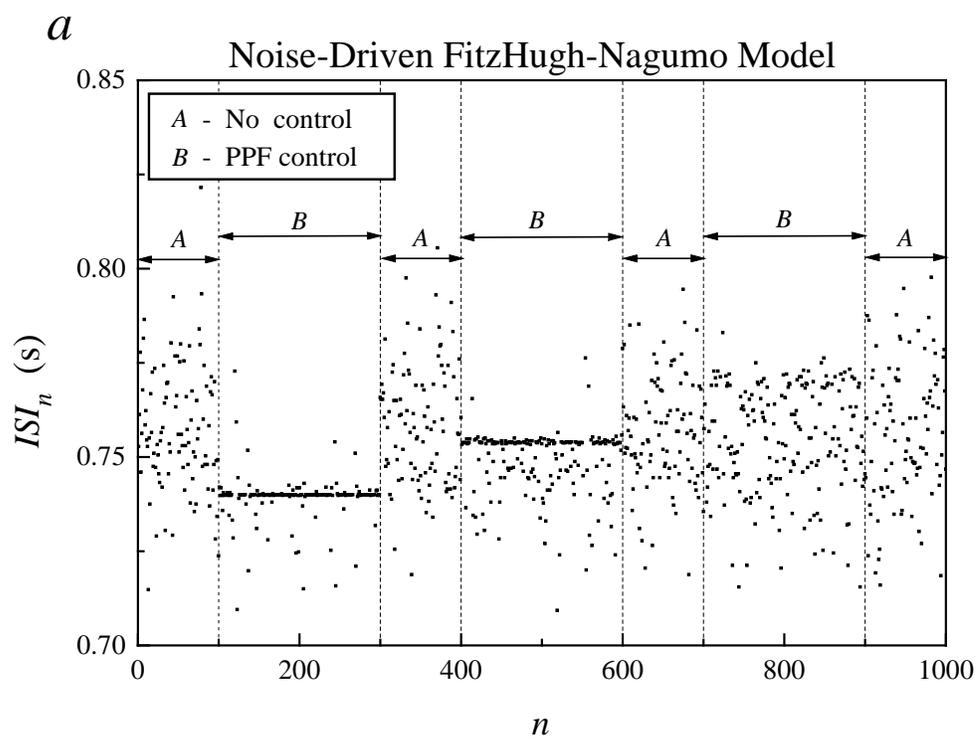

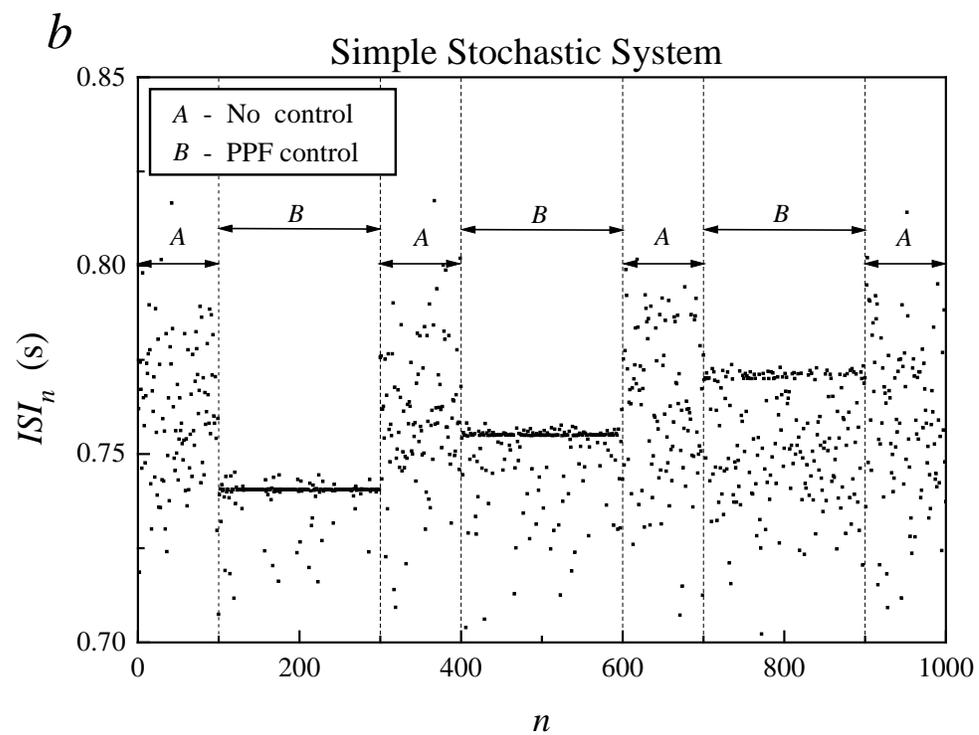

Fig. 3